\documentclass[twocolumn,10pt]{article}

\usepackage[margin=1in]{geometry}

\usepackage{float}
\usepackage{graphicx}
\usepackage{caption}
\usepackage{subcaption}
\usepackage{epsfig} %% for loading postscript figures

\usepackage{authblk}

\usepackage{multirow}

\usepackage{booktabs} % for professional tables

\usepackage{enumitem}

\usepackage{microtype}
\DisableLigatures{encoding = *, family = * }

\newcommand{\gitpage}{https://github.com/PML-UCF/pinn}

\usepackage{amssymb}
\usepackage{amsmath}

\title{Fleet Prognosis with Physics-informed Recurrent Neural Networks}

\author[]{Renato Giorgiani Nascimento\thanks{Graduate Research Assistant}}
\author[]{Felipe A. C. Viana\thanks{Assistant Professor and corresponding author.\\Email: viana@ucf.edu}}
\affil[]{Department of Mechanical and Aerospace Engineering\\
University of Central Florida}
\begin{document}

\maketitle    

%%%%%%%%%%%%%%%%%%%%%%%%%%%%%%%%%%%%%%%%%%%%%%%%%%%%%%%%%%%%%%%%%%%%%%
\begin{abstract}
{
Services and warranties of large fleets of engineering assets is a very profitable business. The success of companies in that area is often related to predictive maintenance driven by advanced analytics. Therefore, accurate modeling, as a way to understand how the complex interactions between operating conditions and component capability define useful life, is key for services profitability. Unfortunately, building prognosis models for large fleets is a daunting task as factors such as duty cycle variation, harsh environments, inadequate maintenance, and problems with mass production can lead to large discrepancies between designed and observed useful lives. This paper introduces a novel physics-informed neural network approach to prognosis by extending recurrent neural networks to cumulative damage models. We propose a new recurrent neural network cell designed to merge physics-informed and data-driven layers. With that, engineers and scientists have the chance to use physics-informed layers to model parts that are well understood (e.g., fatigue crack growth) and use data-driven layers to model parts that are poorly characterized (e.g., internal loads). A simple numerical experiment is used to present the main features of the proposed physics-informed recurrent neural network for damage accumulation. The test problem consist of predicting fatigue crack length for a synthetic fleet of airplanes subject to different mission mixes. The model is trained using full observation inputs (far-field loads) and very limited observation of outputs (crack length at inspection for only a portion of the fleet). The results demonstrate that our proposed hybrid physics-informed recurrent neural network is able to accurately model fatigue crack growth even when the observed distribution of crack length does not match with the (unobservable) fleet distribution.
}
\end{abstract}

%%%%%%%%%%%%%%%%%%%%%%%%%%%%%%%%%%%%%%%%%%%%%%%%%%%%%%%%%%%%%%%%%%%%%%
\section{Introduction}
\label{sec_introduction}
Efficient operation of fleets of engineering assets (e.g., wind turbines, jet engines, etc.) requires the balance between machine performance and maintenance costs (downtime, service, maintenance, repair, and retrofit) \cite{2011_wsc_martin_et_al, 2012_navy_toc}. This has generated a very profitable and competitive market \cite{2017_ge_truechoice, 2017_siemens_services, 2017_lufthansa_engine_services, 2017_gemini_energy_services}. Companies active in this segment base their decisions in models that can predict cumulative distress in critical components as well as machine performance degradation.

\begin{figure}[h]
	\begin{subfigure}{\columnwidth}
		\centering
		\includegraphics[width=0.55\columnwidth]{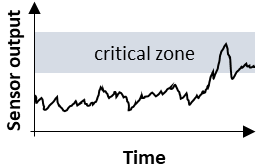}
		\caption{Traditional model for condition-based maintenance.}
		\label{fig_cond_maint}
	\end{subfigure}

	\begin{subfigure}{\columnwidth}
		\centering
		\includegraphics[width=0.55\columnwidth]{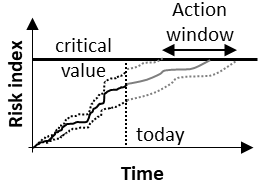}
		\caption{Contemporary model for predictive maintenance.}
		\label{fig_pred_maint}
	\end{subfigure}

	\caption{Comparison of modeling for condition-based and predictive maintenance. While sensor data enables condition-based maintenance, advanced modeling is the one of the main factors to enable predictive maintenance.}
	\label{fig_maintenance}
\end{figure}

These predictive models usually leverage data coming from design, manufacturing, configuration, online sensors, historical records, inspection, maintenance, and awareness network (e.g., location and satellite data). In condition-based maintenance \cite{1995_jqme_tsang, 2006_mssp_jardine_et_al, 2010_ijamt_peng_et_al, 2017_ress_alaswad_xiang}, these models monitor machine status through sensors, as depicted in Fig. \ref{fig_cond_maint} (sensor outputs do not need to be the raw sensor reading and often include simple transformations such as converting time-dependent responses to the frequency domain). While the goal of condition-based maintenance is to perform maintenance only when it is needed, its practical implementation is limited due to dependency on sensors and rule-based models.

Although there is no naming consensus, maintenance driven by sophisticated models (often with an element of machine learning) has been referred to as predictive maintenance. As illustrated in Fig. \ref{fig_pred_maint}, predictive maintenance uses prognosis models to forecast evolution of equipment distress. These prognosis models can be either physics-based or machine learning models (or even a combination of both) \cite{2013_pem_baraldi_et_al, 2015_ress_dawn_et_al}. Without going deeper into that discussion, we believe most practitioners would agree that:
\begin{itemize}[label={\textbullet}]
	\item Machine learning models offer flexibility but tend to require large amounts of data. There is considerable research on the use of traditional and modern machine learning methods for prognosis  \cite{2011_ejor_si_et_al, 2013_ress_tamilselvan_wang, 2013_ress_son_et_al, 2015_ieeetii_susto_et_al, 2018_mssp_khan_yairi, 2018_book_stadelmann_et_al}.
	\item Physics-based models are grounded on first-principles; however, they require good understanding of physics of failure and degradation mechanisms. There is considerable research on the use of physics-based methods for prognosis  \cite{2011_ieee_ac_daigle_goebel, 2013_aiaa_li_et_al, 2017_scitech_ling_et_al, 2019_scitech_yucesan_viana, 2018_ress_iamsumang_et_al}.
	\item The decision between machine learning and physics-based models should be based on factors such as existing knowledge, amount and nature of available data, accuracy requirements, etc.
\end{itemize}

This work is aligned with recent trends of deploying the unparalleled computational power available these days towards building prognosis models that feed of large amounts of data produced by fleets of engineering assets. As we previously mentioned, while research has been done in terms of traditional and modern machine learning methods \cite{2011_ejor_si_et_al, 2013_ress_tamilselvan_wang, 2013_ress_son_et_al, 2015_ieeetii_susto_et_al, 2018_mssp_khan_yairi, 2018_book_stadelmann_et_al}; to the best of our knowledge, there is no work published on physics-informed machine learning methods coined specifically for prognosis. In this contribution, we introduce a novel approach for cumulative damage modeling based on recurrent neural networks. Our formulation is highly conducive for hybrid models (fusing physics-based and machine learning models). We proposed a model of recurrent neural network cell and demonstrate how to apply it in tracking fatigue crack growth at a fleet of engineering assets\footnote{In principle, we do not see why the proposed recurrent neural network cell could not be applied to model other failure mechanisms, such as corrosion, oxidation, etc.}.

The remaining of the paper is organized as follows. Section \ref{sec_rnn_overview} gives an overview on physics-informed neural networks (focusing on recurrent neural networks). Section \ref{sec_cdc_rnn} presents the proposed recurrent neural networks cell and how it is applied to modeling fatigue crack growth. Section \ref{sec_num_exp} describes the case study that we use to illustrate the proposed approach in fleet prognosis. Section \ref{sec_results} presents and discusses the numerical results. Finally, section  \ref{sec_conclusions} closes the paper recapitulating salient points and presenting concluding remarks and future work.

%%%%%%%%%%%%%%%%%%%%%%%%%%%%%%%%%%%%%%%%%%%%%%%%%%%%%%%%%%%%%%%%%%%%%%
\section{Recurrent neural networks physics-informed machine learning}
\label{sec_rnn_overview}
Recurrent neural networks (RNNs) \cite{2016_book_goodfellow_et_al} have been successfully used to model time-series data \cite{1994_ieeetnn_connor_et_al, 2014_isca_sak_et_al, 2015_dsaa_chauban_vig}, speech recognition \cite{2013_icassp_graves_et_al}, text sequence \cite{2011_icml_sutskever_et_al}, and many other sequence modeling applications. In essence, RNNs apply a sequence of transformations to a hidden state in the following fashion:
\begin{equation} \label{eq_rnn_basics}
\mathbf{h}_t = f(\mathbf{h}_{t-1} , \mathbf{x}_t),
\end{equation}
where:
\begin{itemize}[label={\textbullet}]
    \item  $t \in [0,\ldots,T]$ represent the time discretization,
    \item $\mathbf{h} \in \mathbb{R}^{n_h}$ are the states representing the sequence,
    \item $\mathbf{x} \in \mathbb{R}^{n_x}$ are input (and observable) variables, and
    \item $f(.)$ is the transformation to the hidden state.
\end{itemize}

\begin{figure}[h]
	\begin{center}
		\centerline{\includegraphics[width=\columnwidth]{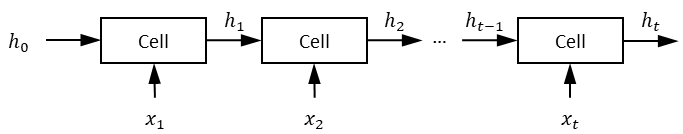}}
		\caption{Unrolled recurrent neural network.}
		\label{fig_rnn_unrolled}
	\end{center}
\end{figure}

As illustrated in Fig. \ref{fig_rnn_unrolled}, the cells of a recurrent neural network repeatedly apply the transformations to the states in the sequence. These states can be observed (i.e., data is collected and available for training of the network) all the time or only at any particular time stamps . This obviously affects the implementation of the loss function used in the training, but it does not affect the implementation of prediction (which only depends on the availability of the inputs).

\begin{figure}[h]
	\centering
	\begin{subfigure}{\columnwidth}
		\centering
		\includegraphics[width=0.85\columnwidth]{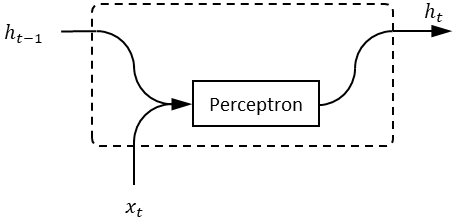}
		\caption{Simple recurrent neural network cell}
		\label{fig_simpleRNN}
	\end{subfigure}

	\begin{subfigure}{\columnwidth}
		\centering
		\includegraphics[width=\columnwidth]{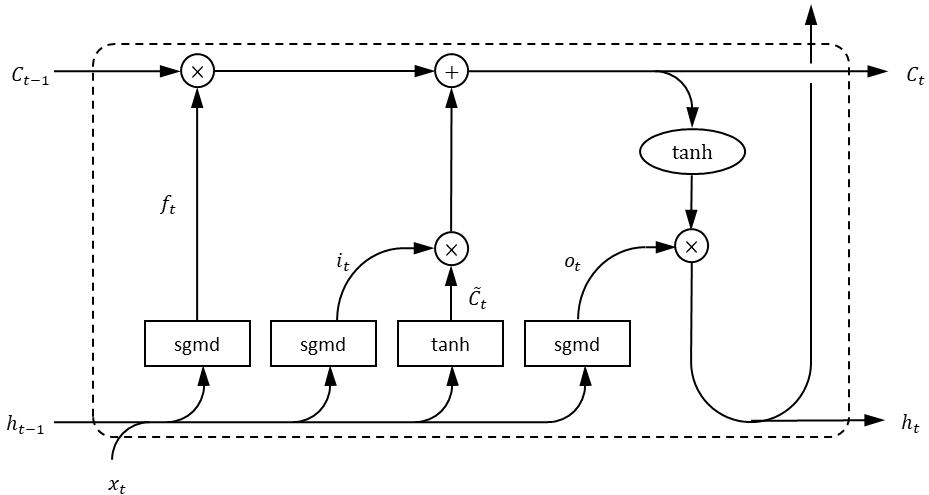}
		\caption{Long short-term memory (LSTM) cell}
		\label{fig_lstm}
	\end{subfigure}

	\caption{Examples of recurrent neural networks cells. In the LSTM cell, the squares are perceptrons with pre-defined activation functions, the oval shape is just the $tanh$ activation.}
	\label{fig_rnn_cells}
\end{figure}

Cells such as the two illustrated in Fig. \ref{fig_rnn_cells} are very common in purely data-driven applications. Fig. \ref{fig_simpleRNN} shows the simplest RNN cell, where a fully-connected dense layer (e.g., a perceptron with a sigmoid activation function) maps the inputs at time $t$ and states at time $t-1$ into the states at time $t$. Figure \ref{fig_lstm} shows an increasingly popular cell architecture, the long short-term memory (LSTM) \cite{1997_nc_hochreiter_schmidhuber}. Without going into further details, the increased complexity of this layer aims at improving the generalization capability of the resulting neural network as well as improving its training (mitigating the vanishing gradient problem \cite{2016_book_goodfellow_et_al}).

How about artificial neural networks based on first principles and designed for engineering applications? Only recently the scientific community has started studying and proposing architectures that leverage formulations based on first principles. In such cases, it is very common that the problem is described by a set of ordinary or partial differential equations. Recent work \cite{2018_jcp_raissi_karniadakis, 2018_nips_chen_et_al, 2018_arXiv_ruthotto_haber} discusses how to essentially use differential equations to constrain the training of feed-forward deep neural networks, multi-layer perceptrons, and RNNs. The interested reader can also find literature on Gaussian processes \cite{2014_nips_schober_et_al, 2018_siamjsc_raissi_et_al}. Nevertheless, to the best of our knowledge, there has not been published work on neural network layers and/or architectures for prognosis.

In this work, we propose a cell for recurrent neural networks inspired on cumulative damage models \cite{1998_ijf_fatemi_yang, 2004_pse_frangopol_el_al}. These models  are often used to describe the irreversible accumulation of damage (progressive distress) throughout the useful life of component or systems. Ultimately, the accumulated damage hits a threshold level that is associated with repair, partial or total replacement, or even worse than that, the retirement, or catastrophic failure of the component or system. From a physics perspective, examples of mechanisms that trigger damage accumulation include: corrosion, erosion, wear, creep, and fatigue, among others.

%%%%%%%%%%%%%%%%%%%%%%%%%%%%%%%%%%%%%%%%%%%%%%%%%%%%%%%%%%%%%%%%%%%%%%
\section{Proposed cumulative damage cell for recurrent neural networks}
\label{sec_cdc_rnn}
\subsection{Base model}
\label{sec_base_rnn}
Cumulative damage models represent damage at time $t$ as an damage increment $\Delta\mathbf{d}_t$ on top of damage $\mathbf{d}_{t-1}$ at previous time step $t-1$
\begin{equation} \label{eq_cumulative_damage}
\mathbf{d}_t = \mathbf{d}_{t-1} + \Delta\mathbf{d}_t,
\end{equation}
where:
\begin{itemize}[label={\textbullet}]
    \item $\mathbf{d}_{t-1}$ is the damage level at time $t-1$, and
    \item  $\Delta\mathbf{d}_t$ is the damage increment, often a function of $\mathbf{d}_{t-1}$ and some other inputs $\mathbf{x}_t$ at time $t$.
\end{itemize}

The detailed modeling of $\Delta\mathbf{d}_t$ depends on the mechanism that governs the physics of failure. In most real applications, the models tend to be built and validated against observed data rather than only using first principles. It is very common to have at least parts of these models coming from curves fitted to a set of controlled experiments. Along the same lines, $\mathbf{x}_t$ is highly application dependent and models for $\mathbf{x}_t$ can become as important as the models for $\Delta\mathbf{d}_t$. For example, the $\Delta\mathbf{d}_t$ model can be built to take the far-field stress $\Delta S_t$ as inputs, which might never be directly observed. In such case, an accurate model for $\Delta S_t$ is just as important as the model for $\Delta\mathbf{d}_t$. This model might come from an engineering model (e.g., using finite element analysis) fed with observable inputs such as pressures, temperatures, altitude, etc.

\begin{figure}[h]
	\begin{center}
		\centerline{\includegraphics[width=0.85\columnwidth]{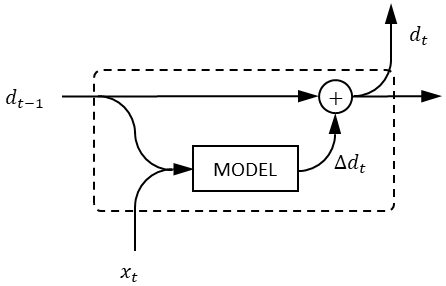}}
		\caption{Cumulative damage cell.}
		\label{fig_cdcell}
	\end{center}
\end{figure}

In the present contribution, we propose the repeating cell illustrated in Fig. \ref{fig_cdcell} to be used while modeling cumulative damage through recurrent neural networks. The ``MODEL'' block maps the inputs $\mathbf{x}_t$ and previous damage $\mathbf{d}_{t-1}$ into a damage increment $\Delta\mathbf{d}_t$. Essentially, ``MODEL'' is the implementation of the physics of failure, which is highly application dependent. Nevertheless, as far as modeling approach, the ``MODEL'' block could be:
\begin{itemize}[label={\textbullet}]
    \item a data-driven model, such as a multi-layer perceptron,
    \item a physics-informed model, with appropriate physical fidelity to reflect the failure mechanism within the expected computational efficiency, or more interestingly,
    \item a hybrid model, where some parts are physics-based while others are data-driven.
\end{itemize}

The decision among these different options depends on the application. Obviously, using an artificial neural network (e.g., pure multi-layer perceptron) as the model in Fig. \ref{fig_cdcell} is a valid approach. However, as we discuss in the next section, a potentially more powerful approach is to build a hybrid model. This can be built with a data-driven component that models applied loads (stresses, temperatures, pressures, etc.) together with a physics-based component that model the mechanism of damage accumulation (e.g., fatigue crack damage accumulation).

In the next section, we discuss how our proposed a recurrent neural network cell can be implemented for fatigue crack growth modeling. As it might be clear  at this point, the proposed recurrent neural network cell could very well be applied to model other failure mechanisms, such as corrosion, oxidation, etc. For convenience, we decided to focus on fatigue crack growth in this contribution.

%%%%%%%%%%%%%%%%%%%%%%%%%%%%%%%%%%%%%%%%%%%%%%%%%%%%%%%%%%%%%%%%%%%%%%
\subsection{Fatigue crack damage accumulation}
\label{sec_fatigue_cda}

From a physics of failure standpoint, fatigue damage can characterized by initiation and propagation of a crack due to cyclic loading \cite{2012_mbm_dowling}. Fatigue crack propagation is usually modeled through Paris's law \cite{1963_jbe_paris}. Mathematically, the crack length $a$ is modeled through an ordinary differential equation that depends on material properties through parameters $C$ and $m$ and loads through the stress intensity range $\Delta K$: 
\begin{equation} \label{eq_paris_law}
\frac{da}{dt} = C \Delta K^m,
\end{equation}
or in its discrete form:
\begin{equation} \label{eq_paris_law_discrete}
a_t  = a_{t-1} + C \Delta K_t^m,
\end{equation}
where $\Delta K_t$ is a function of factors such as localized geometry, previous crack size $a_{t-1}$, and far-field cyclic stress $\Delta S_t$.

In engineering applications (for example, health and reliability management of industrial assets  such as wind turbines, jet engines, etc.), the cyclic loads are either measured or estimated. Then, engineering models map the cyclic loads and current crack length into a stress intensity range. For example, assuming that
a through-the-thickness center crack exists in an infinite plate under the mode I loading condition\footnote{In mode I, load is applied perpendicularly to the crack plane.} and that the far-field cyclic stresses $\Delta S_t$ are available\footnote{As we mentioned before, building accurate estimates of $\Delta S_t$ might just as challenging as modeling damage accumulation itself. Most of the time, $\Delta S_t$ is not measured directly, but instead, it is obtained with the help of some sort of engineering analysis (e.g., through finite element modeling). Even if the instantaneous far-field stresses are available, converting far-field stress time histories into far-field cyclic stresses is usually done through cycle counting approaches such as the rainflow method \cite{2017_astm_cycle_count_stdrd}. Not surprisingly, the cycle counting approaches are application/industry dependent. Even though this is an interesting topic, we consider that the discussion on how to obtain $\Delta S_t$ is outside the scope of this paper.}, the stress intensity range $\Delta K_t $ can be expressed as:
\begin{equation} \label{eq_stress_intensity_range}
\Delta K_t = F\Delta S_t \sqrt{\pi a_{t-1}}.
\end{equation}
where $F$ is a factor that depends on geometry .

\begin{figure}[h]
	\centering
	\begin{subfigure}{0.95\columnwidth}
		\includegraphics[width=\columnwidth]{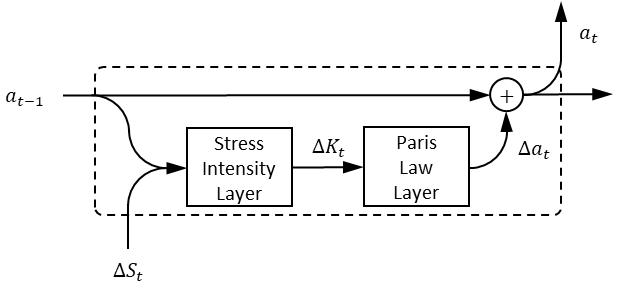}
		\caption{Fully physics-informed cell.}
		\label{fig_crackgrowth_physics}
	\end{subfigure}

	\begin{subfigure}{0.95\columnwidth}
		\includegraphics[width=\columnwidth]{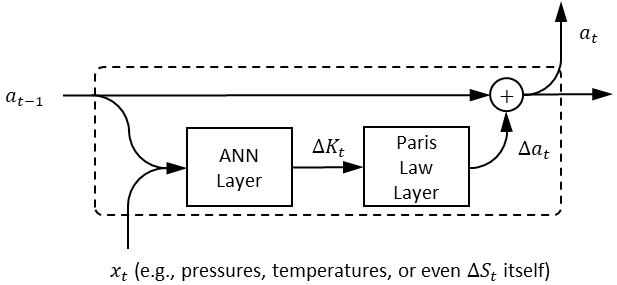}
		\caption{Hybrid physics-informed neural network cell.}
		\label{fig_crackgrowth_mlp}
	\end{subfigure}

	\caption{Examples of crack growth cells for recurrent neural networks. The stress intensity range layer implements $\Delta K_t = F\Delta S_t \sqrt{\pi a_{t-1}}$ and the Paris' law layer implements $\Delta a_t = C \Delta K_t^m$. ANN stands for artificial neural network.}
	\label{fig_crackgrowth_cells}
\end{figure}

This way, Fig. \ref{fig_crackgrowth_physics} illustrates the fully physics-informed implementation of the cumulative damage cell (Fig. \ref{fig_cdcell}). In other words, Fig. \ref{fig_crackgrowth_physics} illustrates how to represent the model described by Eq. \ref{eq_paris_law_discrete} through a recursive neural network. Some interesting observations:
\begin{itemize}[label={\textbullet}]
	\item Stress intensity layer:
		\begin{itemize}[label=-]
			\item implements $\Delta K_t = F\Delta S_t \sqrt{\pi a_{t-1}}$, 
			\item $\Delta S_t$ is the input and $a_{t-1}$ is the state, and
			\item $F$, the geometry factor, can be obtained through engineering analysis. Or alternatively, it can be implemented as trainable parameter (estimated during training of the recurrent neural network).
		\end{itemize}
	\item Paris law layer:
		\begin{itemize}[label=-]
			\item implements  $\Delta a_t = C \Delta K_t^m$, 
			\item $\Delta K_t$ is the input for this layer and comes from a previous layer (physics-based or data driven) that outputs the stress intensity range, and
			\item $C$ and $m$, the Paris' law coefficients, which can be obtained through coupon data\footnote{Many engineering materials have constants documented in handbooks such as \cite{2017_mmpds}.}. Or alternatively, they can be implemented as trainable parameters (estimated in the recurrent neural network training).
		\end{itemize}
\end{itemize}

Another interesting implementation of the base model (Fig. \ref{fig_cdcell}) is the hybrid physics-informed neural network model illustrated in  Fig. \ref{fig_crackgrowth_mlp}. In this case, an artificial neural network layer substitutes the stress intensity range layer. This is very powerful  in certain real life applications, where it might difficult to model $\Delta K_t$ as a function of observed inputs $\mathbf{x}_t$ (such as pressures and temperatures). In such cases, the artificial neural network layer works as a powerful transfer function that maps observed inputs into $\Delta K_t$.

%%%%%%%%%%%%%%%%%%%%%%%%%%%%%%%%%%%%%%%%%%%%%%%%%%%%%%%%%%%%%%%%%%%%%%
\section{Numerical experiments}
\label{sec_num_exp}

\subsection{Case study}

\begin{figure}[h]
	\centering
	\begin{subfigure}{\columnwidth}
		\centering
		\includegraphics[width=0.6\columnwidth]{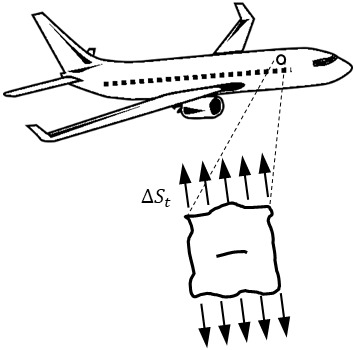}
		\caption{Control point on the aircraft fuselage.}
		\label{fig_aircraft_01}
	\end{subfigure}

	\begin{subfigure}{\columnwidth}
		\centering
		\includegraphics[width=0.8\columnwidth]{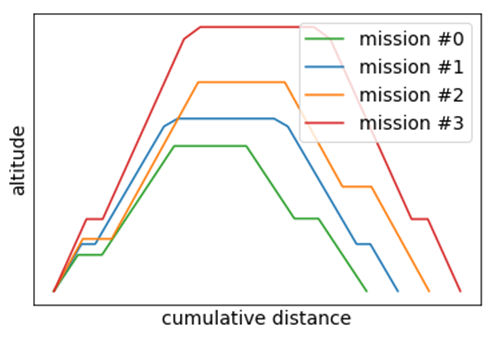}
		\caption{Flight profile for different missions.}
		\label{fig_flight_profile}
	\end{subfigure}

	\begin{subfigure}{\columnwidth}
		\centering
		\includegraphics[width=0.9\columnwidth]{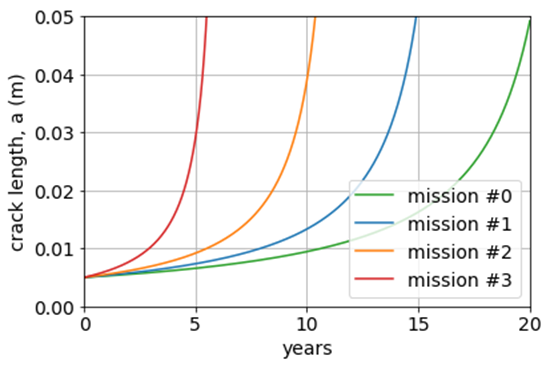}
		\caption{Cumulative damage over time.}
		\label{fig_a_vs_t_by_mission}
	\end{subfigure}

	\caption{Cumulative damage over time for a control point on the aircraft fuselage as a function of mission profile (assuming aircraft consistently flies four missions per day).}
	\label{fig_aircraft_missions}
\end{figure}

Consider a hypothetical control point on an airplane fuselage as illustrated in Fig. \ref{fig_aircraft_01}. For simplicity, assume this airplane was designed to fly four different missions, as shown in Fig. \ref{fig_flight_profile}). As illustrated in Fig. \ref{fig_a_vs_t_by_mission}, fatigue damage accumulates throughout the airplane useful life, following Eq. \ref{eq_paris_law} in propagation. Now, also consider that:

\begin{itemize}[label={\textbullet}]
	\item An original equipment manufacturer, airline company or service provider maintains a large aircraft fleet (hundreds to thousands). Here, we consider a fleet of 300 airplanes.
	\item Airline companies rotate their aircraft fleet through different routes following specific mission mixes. A mission mix determines the proportion of flights from each flight route over the useful life of the aircraft. In this study, we consider the mission mixes detailed in Tab. \ref{tbl_mission_mixes}, with 100 airplanes allocated to each mission mix. For simplicity, each airplane is assigned a fixed percentage flights for each mission that composes the mission mix. These percentages vary uniformly from $0\%$ to $100\%$. Therefore, within the fleet flying mission mix $\#0$, for example, there is one airplane flying $0\%$ of mission $\#0$ and $100\%$ of mission $\#3$, there is another airplane flying $1\%$ of mission $\#0$ and $99\%$ of mission $\#3$, there is yet another airplane flying $2\%$ of mission $\#0$ and $98\%$ of mission $\#3$, an so forth. The same logic applies to the other mission mixes. This way, we synthetically created data for a fleet of 300 airplanes (the missions that each airplane fly over time are drawn from their respective mission mixes, at random). As explained in Section \ref{sec_replication_of_results}, this data is publicly available on GitHub at \gitpage.
	\item Inspection of control points is part of the schedule maintenance activities. Here, we arbitrarily consider that inspection data is available for part of the fleet (60 airplanes out of the 300) at the 5th year.
	\item The initial and the maximum allowable crack lengths are $a_0 = 0.005$ (m) and $a_{max} = 0.05$ (m), respectively.
	\item The metal alloy is characterized by the following Paris' law constants: $C = 1.5\times10^{-11}$, $m = 3.8$.
\end{itemize}

\begin{table}[h]
\caption{Stress range per mission and missions within a mission mix. Percentage flights for each mission vary from $0\%$ to $100\%$.}
	\begin{center}
	\label{tbl_mission_mixes}
	\begin{tabular}{c c c c c}
		\hline
		       & \multicolumn{4}{c}{Mission} \\
                              & \#0     & \#1    & \#2     & \#3    \\
		  $\Delta S$ (MPa) &92.5     & 100     & 110    & 130      \\
                \hline
		\multirow{2}{*}{Mission mix} & \multicolumn{4}{c}{Mission} \\
                              & \#0     & \#1    & \#2     & \#3    \\
                \hline
			   \#0 & \checkmark     &       &         & \checkmark   \\
			   \#1 &       & \checkmark     & \checkmark       &     \\
			   \#2 &       & \checkmark     &         & \checkmark   \\
                \hline
	\end{tabular}
	\end{center}
\end{table}

\begin{figure}[h]
	\centering
	\begin{subfigure}{0.95\columnwidth}
		\includegraphics[width=\columnwidth]{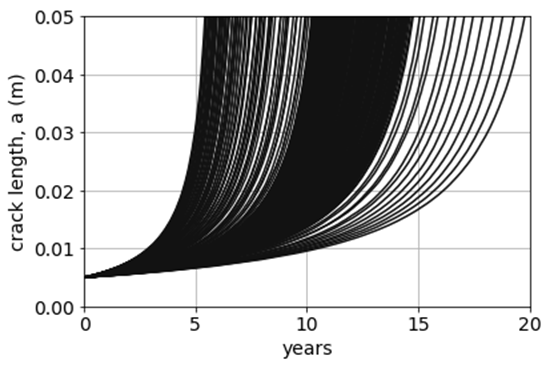}
		\caption{Fatigue crack damage.}
		\label{fig_aTrue}
	\end{subfigure}

	\begin{subfigure}{0.95\columnwidth}
		\includegraphics[width=\columnwidth]{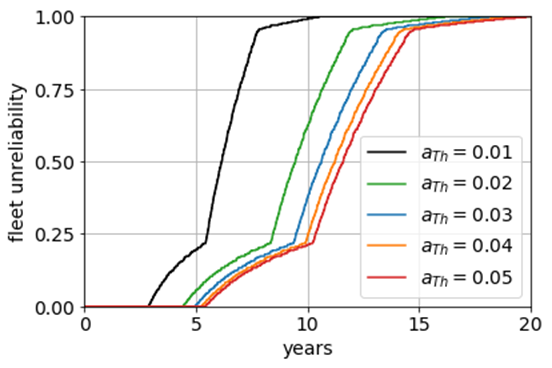}
		\caption{Fleet unreliability over time.}
		\label{fig_fleet_unreliability_true}
	\end{subfigure}

	\caption{Synthetic data used throughout numerical experiments. Fleet unreliability is the proportion of the fleet with fatigue crack length above a threshold $a_{Th}$.}
	\label{fig_synthetic_data}
\end{figure}

Figure \ref{fig_synthetic_data} illustrates how fatigue crack damage accumulates throughout the fleet of 300 airplanes. Figure \ref{fig_aTrue} shows the crack length history over 20 years (some airplanes reach the maximum allowable crack length at approximately 5 years). Figure \ref{fig_fleet_unreliability_true} helps understanding the spread in useful lives throughout the fleet by showing the distribution of time to reach a threshold on crack length.

\subsection{Physics-informed neural network design}

Here for simplicity, we considered the following information is available:
\begin{itemize}[label={\textbullet}]
	\item for every airplane in the fleet: far-field stresses for every flown mission (inputs for the fatigue crack damage accumulation model described in Section \ref{sec_fatigue_cda}), and
	\item for part of the fleet: fatigue crack is observed on 60 airplanes out of the 300 at the end of the 5th year.
\end{itemize}

With that information, we proceed to build a hybrid physics-informed neural network model for fatigue crack damage accumulation. In this model, the stress intensity layer is a multi-layer perceptron and the Paris law layer is physics-based \footnote{It is important to notice that the fully physics-informed model, in which both stress intensity and Paris law layers are physics-based, is used to generate the synthetic data.}. Table \ref{tbl_dklayer} details the multi-layer perceptron used in this work. We decided to use this architecture to illustrate the ability to fit a neural network with a large number of trainable parameters. No attempt was made to further simplify the multi-layer perceptron. In practical applications, we believe reducing the model is worth pursuing, as it could potentially lead to a more manageable number of trainable parameters without sacrificing accuracy.

\begin{table}[H]
\caption{Stress intensity range layer details. Dense $\#0$ is a perceptron without activation function that only scales the inputs (weights and bias selected such that inputs are zero approximately mean and unit standard deviation). Dense $\#1$ to $\#4$  are regular perceptrons with a sigmoid activation functions. Dense $\#5$ is a perceptron without activation function. Finally, PReLU is a parametric rectified linear unit layer (it implements $f(x) = \alpha x$ for $x < 0$, $f(x) = x$ for $ x \geq 0$, where $\alpha$ is a learned array with the same shape as x. This way, the total parameters number of parameters is 1,218 (out of which 1,212 are trainable and 6 are non-trainable ).}
	\begin{center}
	\label{tbl_dklayer}
	\begin{tabular}{llll}
		\hline
		Layer                & Output                        & $\#$ params                                                                                  & Trainable? \\
		\hline
		Dense \#0                 & (None, 2)                        & 6                                                                                  & N                             \\
		Dense \#1                 & (None, 40)                       & 120                                                                                & Y                             \\
		Dense \#2                 & (None, 20)                       & 820                                                                                & Y                             \\
		Dense \#3                 & (None, 10)                       & 210                                                                                & Y                             \\
		Dense \#4                 & (None, 5)                        & 55                                                                                 & Y                             \\
		Dense \#5                 & (None, 1)                        & 6                                                                                  & Y                             \\
		PReLU                     & (None, 1)                        & 1                                                                                  & Y                            \\
		\hline
	\end{tabular}
	\end{center}
\end{table}

Table \ref{tbl_pinncell} further details the hybrid physics-informed recurrent neural network cell we designed. The ``Sequential'' layer is indeed the multi-layer perceptrion (see Tab. \ref{tbl_dklayer}) that estimates the stress intensity range given the far-field stresses and the current fatigue crack length as inputs. The ``ParisLaw'' layer is physics-based and takes the estimated stress intensity range from the ``Sequential layer'' and returns an increment in damage  $\Delta a_t$ (see Section \ref{sec_fatigue_cda}).

\begin{table}[H]
\caption{Physics-informed recurrent neural network cell details. The Sequential layer is the multi-layer perceptron defined in Tab. \ref{tbl_dklayer}. ParisLaw implements the Paris law layer (as detailed in section \ref{sec_fatigue_cda}). Since, the material properties are assumed to be known, the ParisLaw layer has no trainable parameters. This way, the total parameters is 1,220 (out of which 1,212 are trainable and 8 are non-trainable).}
	\begin{center}
	\label{tbl_pinncell}
	\begin{tabular}{llll}
		\hline
		Layer           & Output   & $\#$ params                                                                                  & Trainable? \\
		\hline
		Sequential    & (None, 1)        & 1218                                                                                 & Y (1212)          \\
		ParisLaw      & (None, 1)        & 2                                                                                  & N                            \\
		\hline
	\end{tabular}
	\end{center}
\end{table}

Our implementation is all done in TensorFlow using the Python application programming interface (version 1.11.0)\footnote{www.tensorflow.org.}. Some of the TensorFlow nomenclature is  used here (e.g., we use ``Dense'' layer when we refer to a perceptron). The ``ParisLaw'' layer is also implemented in TensorFlow by extendig the ``tf.keras.layers.Layer'' class.

\subsection{Replication of results}\label{sec_replication_of_results}
Data and codes (including our implementation for both the multi-layer perceptron,  the stress intensity and Paris law layers, the cumulative damage cell, as well as python driver scripts) used in this manuscript are publicly available on GitHub at \gitpage. The data and code are released under the MIT License.

%%%%%%%%%%%%%%%%%%%%%%%%%%%%%%%%%%%%%%%%%%%%%%%%%%%%%%%%%%%%%%%%%%%%%%
\section{Results and discussion}
\label{sec_results}
\begin{figure}[h]
	\centering
	\begin{subfigure}{\columnwidth}
		\centering
		\includegraphics[width=0.95\columnwidth]{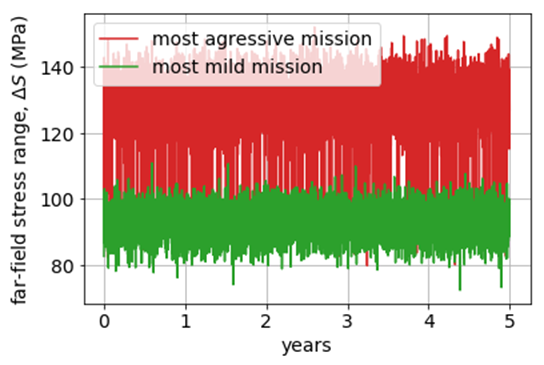}
		\caption{Example of two missions histories of airplanes with observed crack length at the 5th year.}
		\label{fig_mission_examples}
	\end{subfigure}

	\begin{subfigure}{\columnwidth}
		\centering
		\includegraphics[width=0.95\columnwidth]{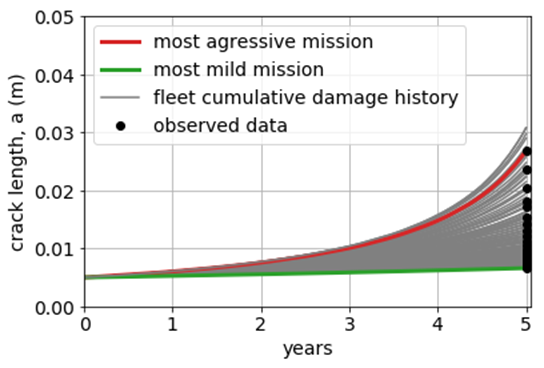}
		\caption{Fatigue crack length history for the fleet and observations at the 5th year.}
		\label{fig_observed_damage}
	\end{subfigure}

	\begin{subfigure}{\columnwidth}
		\centering
		\includegraphics[width=0.95\columnwidth]{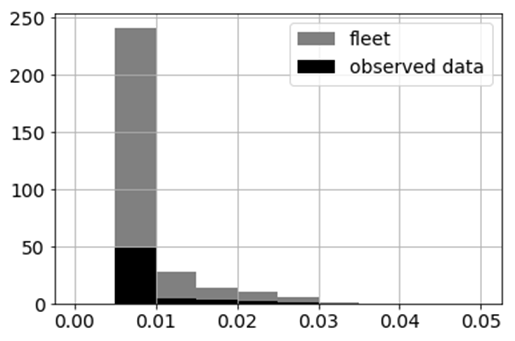}
		\caption{Histograms of fatigue crack length at the 5th year.}
		\label{fig_a5yrs_fleet_obs}
	\end{subfigure}

	\caption{Synthetic data used throughout numerical experiments.}
	\label{fig_observed_data}
\end{figure}

\begin{figure}[h]
	\begin{center}
		\centerline{\includegraphics[width=0.99\columnwidth]{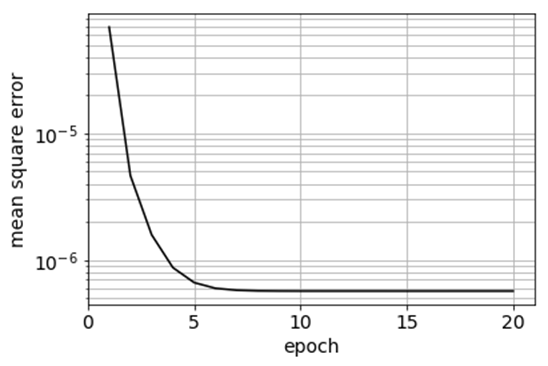}}
		\caption{Mean square error through training.}
		\label{fig_loss_history}
	\end{center}
\end{figure}

Figure \ref{fig_observed_data} illustrates the data used for training and later assessing the prediction accuracy of the hybrid physics-informed neural network model. Figure \ref{fig_mission_examples} shows the complete mission history in terms of far-field stresses for two airplanes that will have damage data collected at the 5th year inspection. The consistent difference in stress range amplitude leads to significant difference in damage accumulation. This is clear in in Fig. \ref{fig_observed_damage}, which also brings the crack length history for the entire fleet as well as highlights the 60 observed crack lengths at the end of the 5th year. Note that:
\begin{itemize}[label={\textbullet}]
	\item for this particular example, the observed data is a fair subsample of the entire fleet, as shown in Fig. \ref{fig_a5yrs_fleet_obs},
	\item while our hybrid physics-informed recurrent neural network is capable of predicting the entire crack growth history; during its training, only crack length observed at the 5th year is used in the loss function, and
	\item for this problem, the mean square error converges fast throughout the training, as shown in Fig. \ref{fig_loss_history}.
	\item we do not report computational cost here since it is actually very small (training and prediction in only few minutes in a very modest laptop computer).
\end{itemize}

\begin{figure}[h]
	\centering
	\begin{subfigure}{\columnwidth}
		\centering
		\includegraphics[width=0.99\columnwidth]{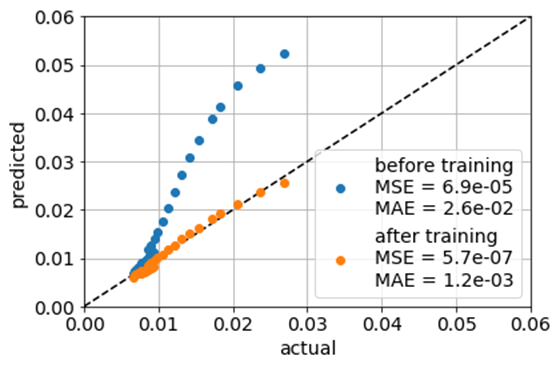}
		\caption{Predictions at the training data (60 airplanes).}
		\label{fig_prediction_at_training}
	\end{subfigure}

	\begin{subfigure}{\columnwidth}
		\centering
		\includegraphics[width=0.95\columnwidth]{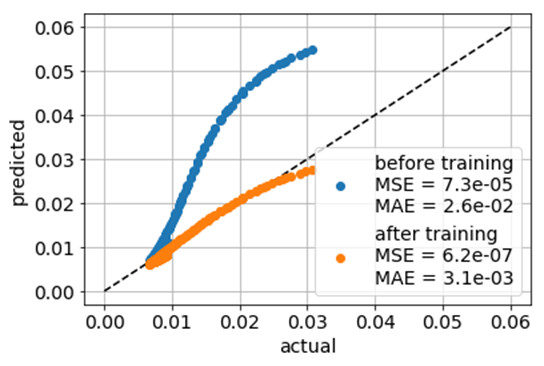}
		\caption{Predictions at the entire fleet (300 airplanes).}
		\label{fig_prediction_at_fleet}
	\end{subfigure}

	\caption{Predictions before and after training. MSE and MAE stand for mean squared error and maximum absolute square error, respectively.}
	\label{fig_predictions}
\end{figure}

Figure \ref{fig_predictions} shows the predictions at the 5th year before and after training of the neural network. While Fig. \ref{fig_prediction_at_training} shows only the predictions at the training set (60 airplanes), Fig. \ref{fig_prediction_at_fleet} shows the predictions at the entire fleet (300 airplanes). In both cases, the model initially tends to over-predict the large crack lengths. After the neural network is trained, the predictions are in good agreement with the actual values.

\begin{figure}[h]
	\centering
	\begin{subfigure}{\columnwidth}
		\centering
		\includegraphics[width=0.95\columnwidth]{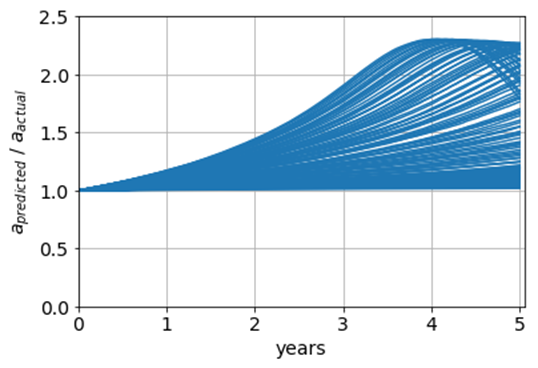}
		\caption{Before training.}
		\label{fig_aPred_over_aActual_before}
	\end{subfigure}

	\begin{subfigure}{\columnwidth}
		\centering
		\includegraphics[width=0.95\columnwidth]{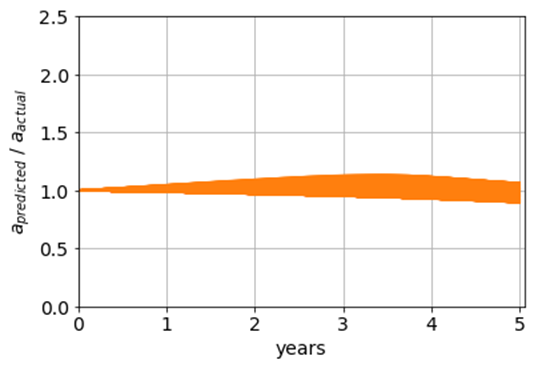}
		\caption{After training.}
		\label{fig_aPred_over_aActual_after}
	\end{subfigure}

	\caption{Ratio between predicted and actual crack length for the entire fleet (300 airplanes).}
	\label{fig_aPred_over_aActual}
\end{figure}

Figure \ref{fig_aPred_over_aActual} shows the ratio between the predicted and actual crack growth over time for the entire fleet. The ratio is illustrated for the model predictions before the training (Fig. \ref{fig_aPred_over_aActual_before}), as well as after the training (Fig. \ref{fig_aPred_over_aActual_after}). Initially, the model is biased (i.e., over-predicts large crack lenghts) and the predictions are within a $0\%$ to $+130\%$ range of the actual crack length. After the it is trained with the 60 observations, the model becomes unbiased and predictions always stay within $\pm15\%$ of the actual crack length.

\begin{figure}[h]
	\centering
	\begin{subfigure}{\columnwidth}
		\centering
		\includegraphics[width=0.95\columnwidth]{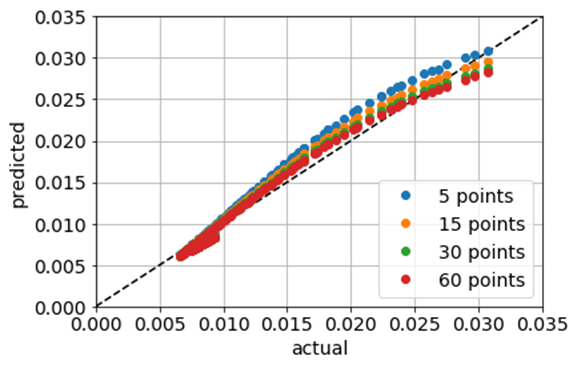}
		\caption{Predictions versus actual crack length.}
		\label{fig_number_pred_actual}
	\end{subfigure}

	\begin{subfigure}{\columnwidth}
		\centering
		\includegraphics[width=0.95\columnwidth]{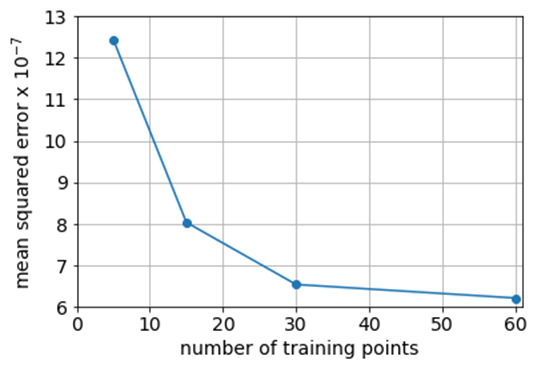}
		\caption{Mean squared error versus number of training points.}
		\label{fig_number_points_mse}
	\end{subfigure}

	\caption{Effect of number of training points in crack length predictions for entire fleet (300 airplanes).}
	\label{fig_number_points}
\end{figure}

Figure \ref{fig_number_points} illustrates how the number of training data affects the prediction accuracy of the resulting physics-informed neural network. Again, besides time series for loads (inputs), only observations for crack at the 5th year are used for training the model. Figure \ref{fig_number_pred_actual} shows the predictions versus actual crack lengths for the entire fleet (300 airplanes) at the end of the 5th year. Incredibly, even with as little as 5 observations (full load histories and crack length at the 5th year), the model is capable of producing relatively accurate predictions. Figure \ref{fig_number_points_mse} shows the convergence of mean squared error as a function of number of training points. One can think that the more training points, the more accurate the resulting model becomes. Nevertheless, in this application, a diminishing return seems to exist after 30 points.

Modern neural network architectures are usually associated with data-intensive applications. One might be tempted to think that the application shown in this paper do not conform with that. After all, the study only used 5 to 60 observations. This only means that there are 5 to 60 output observations. However, at a rate of 4 flights per day, in a period of 5 years, this means that we observed 5 to 60 time histories of 7,300 data points each (total of 36,500 to 438,000 data points). The judgment of whether or not this constitutes a large data set is outside the scope of this paper. Nevertheless, this seems to indicate that the physics of failure (through Paris law) imposes a strong constraint in this problem, helping training the neural network (by virtually filtering/guiding which crack growth paths are physically plausible).

\begin{figure}[h]
	\begin{subfigure}{\columnwidth}
		\centering
		\includegraphics[width=0.95\columnwidth]{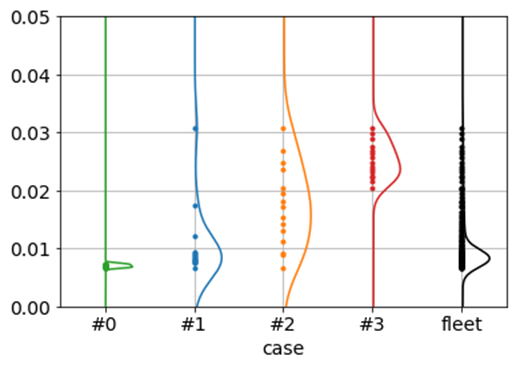}
		\caption{Distribution of observed (colored) and fleet (black) crack length.}
		\label{fig_unbalanced_hist}
	\end{subfigure}
	
	\centering
	\begin{subfigure}{\columnwidth}
		\centering
		\includegraphics[width=0.99\columnwidth]{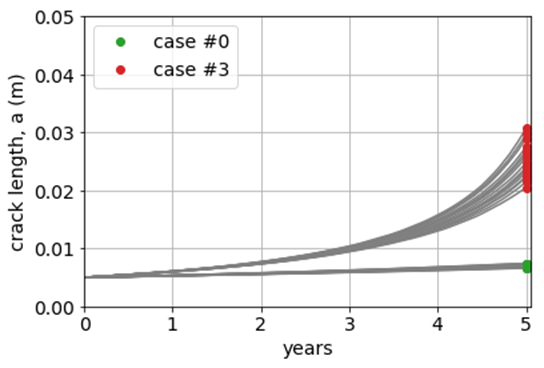}
		\caption{Unobserved fatigue crack length history (gray) and observed data (colored).}
		\label{fig_unbalanced_damage}
	\end{subfigure}
	
	\begin{subfigure}{\columnwidth}
		\centering
		\includegraphics[width=0.99\columnwidth]{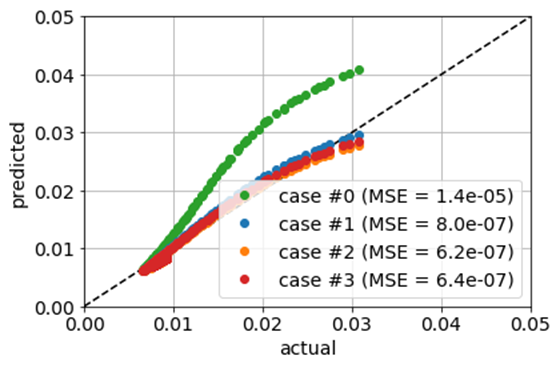}
		\caption{Predictions versus actual crack length at the 5th year.}
		\label{fig_unbalanced_pred_actual}
	\end{subfigure}

	\caption{Effect of distribution of training data (15 points) in crack length predictions for entire fleet (300 airplanes).}
	\label{fig_unbalanced}
\end{figure}

Last but not least, we also studied the effect of the distribution of crack length observations used for training the recursive neural network. For the sake of illustration, consider that the training set consists of observations for crack lengths and far-field cyclic stress at 15 different airplanes. One might be interested in looking at how well the resulting model is when the crack length observation is biased towards the low values, or towards high values, or maybe, the distribution of crack length observations does not reflect the fleet distribution. Figure \ref{fig_unbalanced} shows the summary of results for this part of the study. Figures \ref{fig_unbalanced_hist} illustrates the distributions for the four considered cases:
\begin{itemize}[label={\textbullet}]
	\item Case $\#0$: observed crack length distribution biased towards low values,
	\item Case $\#1$: observed crack length distribution approximates the fleet distribution (this is the same data used to generate the previous results, see Fig. \ref{fig_a5yrs_fleet_obs}),
	\item Case $\#2$: observed crack length distribution has considerably larger spread when compared to the fleet, and
	\item Case $\#3$: observed crack length distribution biased towards high values.
\end{itemize}

As shown in Fig. \ref{fig_unbalanced_damage}, the observed crack length at the 5th year implies in very different damage accumulation histories (due to different far-field cyclic stress histories). Figure \ref{fig_unbalanced_pred_actual} shows how the prediction results compare against fleet data at the 5th year. Interestingly, except for case $\#0$, the trained physics-informed recurrent neural network was able to predict crack length (there is only minor differences among cases $\#1$, $\#2$, and $\#3$). This indicates that:
\begin{itemize}[label={\textbullet}]
	\item as long as the range of observed crack length covers the plausible crack lengths at the fleet level, the resulting tends to be accurate, and
	\item the distribution of observed crack length has minor effects on the resulting network (as long as the range covers plausible fleet outcomes).
\end{itemize}

%%%%%%%%%%%%%%%%%%%%%%%%%%%%%%%%%%%%%%%%%%%%%%%%%%%%%%%%%%%%%%%%%%%%%%
\section{Conclusions and future work}
\label{sec_conclusions}

In this contribution, we proposed a novel cell to be used while modeling cumulative damage through recurrent neural networks. This proposed cell is designed such that cumulative damage models can be built using purely data-driven layers (such as the conventional multi-layer perceptrons), purely physics-based layers (such as the stress intensity and Paris law layers used here to generate fatigue crack data), or more interestingly, hybrids of physics-informed and data-driven layers (as the model discussed in this paper). Then, we designed a simple numerical experiment where:

\begin{itemize}[label={\textbullet}]
	\item a physics-informed recurrent neural network is used to model fatigue crack growth on a fleet of 300 airplanes;
	\item the physics-informed recurrent neural network is built with a multi-layer perceptron that models the stress intensity range and feeds into a Paris law layer (which is physics-informed);
	\item the parameters for the multi-layer perceptron are trainable and the ones for the Paris law layer are fixed; and
	\item airplanes in the fleet are subjected to three different mission mixes (controlling the severity of each flight).
\end{itemize}

With the help of this numerical study:
\begin{itemize}[label={\textbullet}]
	\item We demonstrated that the proposed physics-informed neural network cell for cumulative damage successfully models fatigue crack growth;
	\item We studied the effect of the number of training data in the accuracy of the crack length predictions at the fleet level after five years worth of operation. We learned that even with reduced number of data points, the proposed physics-informed neural network can approximate fatigue crack growth. Obviously, we observed that the more training points are used, the more accurate the model becomes (although there seems to exist a limit).
	\item We studied the effect of the training set balance (in terms of coverage of the observed crack length as compared with the fleet status). We learned that results are highly sensitive to the range of the training data with regards to the possible observable output values. Interestingly, it seems that the accuracy of the resulting model is not related to how well the training data reflects the actual output distribution (as long as it covers the range of plausible observations).
\end{itemize}

The results obtained so far are promising, and we want to extend the study and include, among other factors:
\begin{itemize}[label={\textbullet}]
	\item Improved physics of failure models: by including both initiation and propagation in the cumulative damage.
	\item Several sources of uncertainty, such as:
		\begin{itemize}[label=-]
			\item scatter in material properties (i.e., uncertainty in Paris law coefficients),
			\item field inspection, in terms of both damage detection (via probability of detection) and damage quantification (such as measurement error), and
			\item service level (quality of repair after observed damage reaches a pre-defined threshold).
		\end{itemize}
	\item Strategies for services: repair and replacement as performed by operators and service providers.
\end{itemize}

%%%%%%%%%%%%%%%%%%%%%%%%%%%%%%%%%%%%%%%%%%%%%%%%%%%%%%%%%%%%%%%%%%%%%%
\section*{Acknowledgments}
This work was supported by the University of Central Florida (UCF). Nevertheless, any view, opinion, findings and conclusions or recommendations expressed in this material are those of the authors alone. Therefore, UCF does not accept any liability in regard thereto. 

%%%%%%%%%%%%%%%%%%%%%%%%%%%%%%%%%%%%%%%%%%%%%%%%%%%%%%%%%%%%%%%%%%%%%%
\bibliography{pmlBibtex}
\bibliographystyle{ieeetr}

\end{document}